\documentclass[11pt,a4paper]{article}
\usepackage[T1]{fontenc}
\usepackage[utf8]{inputenc}
\usepackage{authblk}

\usepackage{amsmath,amsfonts,amssymb,amsthm,nicefrac,mathrsfs,cancel}    
\usepackage{graphicx}   
\usepackage{natbib}
\usepackage{epstopdf}
\usepackage{subfig}

\DeclareGraphicsExtensions{.pdf,.jpeg,.png}
\graphicspath{{img/}}

\newtheorem{theorem}{Theorem}
\newtheorem{lemma}{Lemma}

\providecommand{\ds}{\displaystyle}
\providecommand{\w}{\ensuremath{\omega} }
\providecommand{\abs}[1]{\ensuremath{\left| #1 \right|}}
\providecommand{\sech}{\text{sech}}
\providecommand{\R}{\ensuremath{\mathbb{R}}}
\providecommand{\ph}{\ensuremath{\Phi}}
\providecommand{\vp}{\ensuremath{\Upsilon}}
\providecommand{\pd}[2]{\ensuremath{\frac{\partial #1}{ \partial #2}}}
\providecommand{\inprod}[1]{\ensuremath{\left\langle#1\right \rangle }}
\providecommand{\intrr}[1]{\ensuremath{\int_{\mathbb{R}^{2}} #1 d \mathbf{x}}}

\providecommand{\cph}{\ensuremath{\ph}}

\providecommand{\vph}{\ensuremath{\varphi}}

\providecommand{\Th}{\ensuremath{ \Theta}}
\newcommand{\divergence}{\ensuremath{\nabla \cdot}}
\providecommand{\divg}{\divergence}

\providecommand{\bo}[1]{\ensuremath{\mathcal{O}\left(#1\right)}}
\providecommand{\bxi}{\ensuremath{\boldsymbol{\xi}}}
\providecommand{\e}{\ensuremath{ \epsilon }}

\providecommand{\lap}{\ensuremath{\nabla^2}}
\providecommand{\grad}{\ensuremath{\nabla}}

\providecommand{\sinth}{\ensuremath{\sech\left(\rho - \frac{1}{\w} \right) }}

\providecommand{\intnco}[1]{\ensuremath{\int_{|\mathbf{x}| < \rho_*} #1 d \mathbf{x}}}

\date{}
\title{Perturbation Theory for Propagating Magnetic Droplet Solitons}
\author[1]{L. D. Bookman}
\author[2]{ M. A. Hoefer}
\affil[1]{Department of Mathematics, North Carolina State
  University, Raleigh, North Carolina 27695, USA}
\affil[2]{Department of Applied Mathematics, University of Colorado, Boulder, Colorado 80309, USA}

\begin{document}
  \maketitle

\begin{abstract}
  Droplet solitons are a strongly nonlinear, inherently dynamic
  structure in the magnetization of ferromagnets, balancing dispersion
  (exchange energy) with focusing nonlinearity (strong perpendicular
  anisotropy).  Large droplet solitons have the approximate form of a
  circular domain wall sustained by precession and, in contrast to
  single magnetic vortices, are predicted to propagate in an extended,
  homogeneous magnetic medium.  In this work, multiscale perturbation
  theory is utilized to develop an analytical framework for
  investigating the impact of additional physical effects on the
  behavior of a propagating droplet. After first developing soliton
  perturbation theory in the general context of Hamiltonian systems, a
  number of physical phenomena of current interest are investigated.
  These include droplet-droplet and droplet-boundary interactions,
  spatial magnetic field inhomogeneities, spin transfer torque induced
  forcing in a nanocontact device, and damping.  Their combined
  effects demonstrate the fundamental mechanisms for a previously
  observed droplet drift instability and under what conditions a
 slowly propagating droplet can be supported by the nanocontact, important
  considerations for applications.  This framework emphasizes the
  particle-like dynamics of the droplet, providing analytically
  tractable and practical predictions for modern experimental
  configurations.
\end{abstract}

\section{Introduction}
\label{sec:introduction}

The ability to excite, probe, and control magnetic media at the
nanometer scale has enabled new applications in spintronics and
magnonics as well as the exploration of new physics.  Of particular,
recent interest is the generation of coherent structures within the
magnetization, a vector field describing the magnetic dipole moments
of a magnetic material.  Coherent structures observed experimentally
include those with nontrivial topology such as magnetic vortices
\cite{uhlir_dynamic_2013,pulecio_coherence_2014} and skyrmions
\cite{nagaosa_topological_2013,fert2013skyrmions,yu_biskyrmion_2014}
as well as nontopological droplet solitons
\cite{mohseni2013spin,mohseni_magnetic_2014,chung_spin_2014,macia2014stable}.
The magnetic droplet soliton (droplet hereafter) is a coherently
precessing, nanometer-scale localized wave structure exhibiting
strongly nonlinear effects \cite{kosevich1990magnetic}.  Its first
observation \cite{mohseni2013spin} was enabled by a nano-contact
spin-torque oscillator (NC-STO) device, which provides the necessary
forcing to oppose magnetic damping, hence has been termed a
dissipative droplet soliton \cite{hoefer2010theory}.

\begin{figure}[htbp]
\centering
\begin{tabular}{cc}
\includegraphics{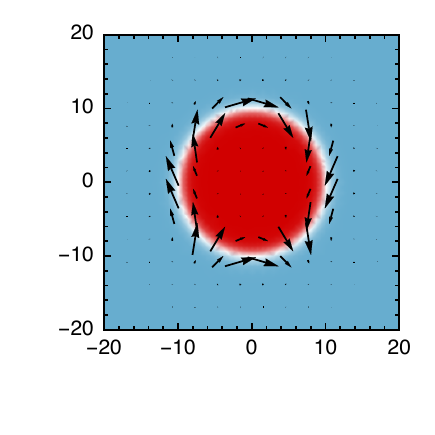} &\includegraphics{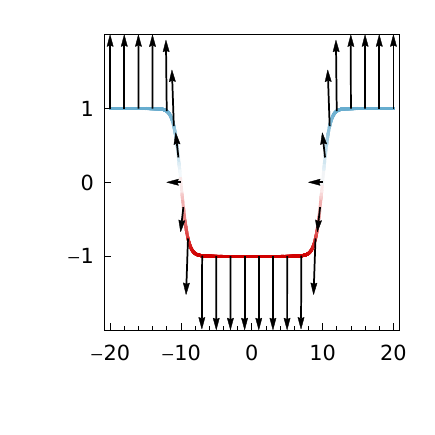} 
\begin{picture}(30,0)
 \put(-10,13.25) {\includegraphics{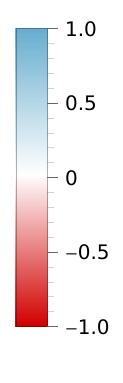}} 
 \put(-162,35){(a)}
 \put(-25,35){(b)}
 \put(-57,15){\scalebox{.75}{$x$}}
  \put(-194.5,15){\scalebox{.75}{$x$}}
    \put(-250,71){\scalebox{.75}{$y$}}
 \put(-115,71){\scalebox{.75}{$m_z$}}
  \put(16,71){\scalebox{.75}{$m_z$}}
 \end{picture}
\end{tabular}
\caption{(a) Contour (color is $m_z$) and vector field, $(m_x,m_y)$,
  plot of an approximate, propagating magnetic droplet, described in
  Sec.~\ref{sec:droplet}, moving to the right with speed $V = 0.02$ and
  precessing with frequency $\w = 0.1$. (b) The magnetization
  evaluated on the line $y=0$ and superimposed upon the $m_z$
  profile. There is no phase winding across this nontopological
  structure.  }
\label{fig:droplet}
\end{figure}
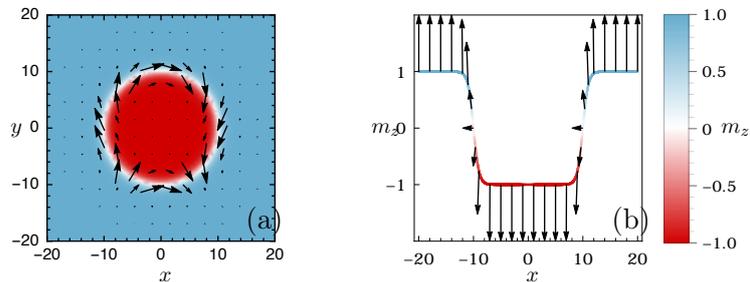

The droplet is theoretically understood as a solution of a
conservative Landau-Lifshitz partial differential equation modeling
spatio-temporal dynamics in an ultra thin, two-dimensional magnetic
film with uniaxial, perpendicular anisotropy.  When the stationary
droplet's precessional frequency is close to the local, magnetic field
induced (Zeeman) frequency, it resembles a circular domain wall of
large radius, almost reversed at its core.  These large droplets can
propagate, coinciding with a superimposed wave structure to the
otherwise spatially homogeneous phase.  An example is shown in
Fig.~\ref{fig:droplet}.

There are a rich variety of physical mechanisms
that can change the orientation of the local dipole moment in a
ferromagnet and hence influence the particle-like behavior of a
droplet.  An example already mentioned is that of the NC-STO and
damping.  This motivated us to develop a droplet soliton perturbation
theory that allows for the analysis of droplet dynamics in the
presence of a large number of physical perturbations
\cite{bookman2013analytical}.  This theory describes soliton dynamics
via a finite dimensional dynamical system representing the adiabatic
evolution of the droplet's four parameters (center, precessional
frequency, and phase) resulting from
perturbations.

While the theory provided fundamental explanations of droplet physics,
it was limited to almost stationary droplets, i.e., droplets of
negligible momentum, which manifested as a constraint equation on the
dynamics.  Furthermore, important physical effects such as droplet
acceleration due to a magnetic field gradient
\cite{kosevich1998magnetic,kosevich2001relaxation,hoefer2012propagation}
and droplet interactions
\cite{piette1998localized, maiden2014attraction} were excluded from the
theory.

Soliton perturbation theory has been successfully used to describe
dynamics in many physical systems
\cite{kivshar1989dynamics,sanchez_collective_1998}, notably solitons
in Nonlinear Schr\"{o}dinger (NLS) type equations modeling, for
example, optical fibers \cite{yang_nonlinear_2010} and Bose-Einstein
condensates \cite{kevrekidis_emergent_2008}.  The central idea is to
project the perturbed PDE dynamics onto the unperturbed soliton
solution manifold, allowing for adiabatic evolution of the soliton's
parameters.  The resulting modulation equations can be obtained in
different ways, of which multiple scales perturbation theory
\cite{keener_solitons_1977} or perturbed conservation laws
\cite{kivshar1989dynamics} are perhaps the most common.  Both
approaches have been shown to be equivalent in some specific
applications, see, e.g., \cite{ablowitz2009asymptotic}, but the
conservation law approach is powerful in its simplicity.  However, it
can be unclear which balance laws to use, especially when higher order
information is sought.  The governing Landau-Lifshitz equation for
magnetization dynamics is a strongly nonlinear, vectorial
generalization of the NLS equation that lacks Galilean invariance.
There exist two-dimensional moving droplet solutions
\cite{piette1998localized,hoefer2012propagating} described by six
independent parameters.  As such, droplet perturbation theory is
significantly more complex than its NLS soliton counterpart, so a
structured approach is desirable.  

The earliest, most general formulation of soliton perturbation theory
we have found in the literature is given in
\cite{keener_solitons_1977}, where multiple scales perturbation theory
and the Green's function for the linearized operator are utilized.  An
alternative, rigorous approach was developed for the modulational
stability of solitons in NLS equations in
\cite{weinstein1985modulational} based on multiple scales and the
generalized nullspace of the linearized operator.  Motivated by the
challenges associated with magnetic droplet soliton perturbation
theory, we revisit the general approach for perturbed Hamiltonian
systems, utilizing multiple scales and the generalized nullspace
formulation.  In Section \ref{sec:hamiltonian_systems}, we provide
necessary conditions and expressions for the resulting modulation
equations. Our approach preserves much of the generality of the Green's 
function approach with the comparative accessibility of Weinstein's approach
to NLS. Additionally, we demonstrate that these methods do generalize to soliton perturbation in higher dimension, 
beyond (1+1)D, and how they can be used to determine the evolution of large numbers of parameters for a single soliton. 

Two-dimensional moving droplet solutions have been computed
\cite{piette1998localized,hoefer2012propagating} and studied
asymptotically in the weakly nonlinear regime \cite{ivanov2001small}.
They can be accelerated and controlled by an inhomogeneous, external
magnetic field \cite{hoefer2012propagation} and exhibit novel
interaction properties \cite{maiden2014attraction}.  Droplets can also
experience a drift instability in NC-STOs, whose origin is not
well-understood \cite{hoefer2010theory}.  Because droplets are
relatively new physical features of nanomagnetic systems, they hold
potential for applications such as spintronic information storage and
transfer or probing material properties.  Moreover, their fundamental
physics are not well understood.  A more general theory to describe
the motion of droplets in realistic physical systems is desirable.

Toward this end, we here present an analytical framework for
investigating the impact of a large class of physical effects on a six
parameter propagating droplet soliton.  In Section \ref{sec:droplet},
we derive an approximate solution for the propagating droplet in the
close to Zeeman frequency (large mass) and small velocity (order one
momentum) regime, which greatly reduces the complexity of the
asymptotic theory developed.  This allows for explicit analytical
results, a powerful feature of soliton perturbation theory considering
the droplet's strongly nonlinear qualities.  The main result of this
paper, the modulation equations, is presented in Section
\ref{sec:general_result}, utilizing the general soliton perturbation
theory formulation from Section \ref{sec:hamiltonian_systems}.  The
versatility of this framework is subsequently discussed in Sections
\ref{sec:applications} and \ref{sec:interacting} through an
investigation of a series of perturbations of current physical
interest.  In particular, we analytically demonstrate a mechanism
leading to the NC-STO droplet drift instability and why a droplet can
be attracted or repelled by a ferromagnetic boundary.

The model under study here is the perturbed Landau-Lifshitz equation
for the magnetization, $\mathbf{m} = [m_x,m_y,m_x]$, of a two-dimensional
ferromagnetic film in nondimensional form
\begin{equation}
  \begin{split}
    \frac{\partial \mathbf{m}}{\partial t} &= -\mathbf{m}\times
    \mathbf{h}_{\rm eff}  + \epsilon \mathbf{p}, \quad \mathbf{m}: \R^2 \times
    (0,\infty) \to \mathbb{S}^2,\\
    \mathbf{h}_{\rm eff}  &= \lap \mathbf{m} +
    \left(h_0 + m_z\right) \mathbf{\hat{z}}, \quad \lim_{|\mathbf{x}|
      \to \infty} \mathbf{m} = \mathbf{\hat{z}},
  \end{split}
\label{eq:nondimensional_ll}
\end{equation}
where $\mathbf{p}$ is a perturbation that preserves the
magnetization's length ($\mathbf{p}\cdot \mathbf{m} \equiv 0$) and
$0<\epsilon\ll 1$ is a small parameter encoding the strength of the
perturbation.  
The perturbation can depend explicitly on time so long as the
variation is slow.  That is, the perturbation depends only on the slow
temporal coordinate $T = \epsilon t$.  The perpendicular, external
magnetic field $h_0 \mathbf{\hat{z}} = h_0(\mathbf{X},T)
\mathbf{\hat{z}}$ can be large and slowly varying in space
($\mathbf{X} = \epsilon \mathbf{x}$) and time.  Further
generalizations to rapidly varying perturbations are possible.  A
necessary requirement for the existence of droplet solitons is strong
perpendicular anisotropy, encoded in the orientation of the $m_z
\mathbf{z}$ term in the effective field $\mathbf{h}_{\rm eff}$.  See
\cite{bookman2013analytical} for the derivation and
nondimensionalization of this model.

\section{Adiabatic Dynamics for Hamiltonian Systems} 
\label{sec:hamiltonian_systems}
The main aim of this paper is to determine the evolution of parameters
of the droplet soliton in response to a fairly general class of
perturbations. The equations determining the time-evolution of
parameters will be referred to as modulation equations.  Before
performing that analysis directly, we first consider the more general
context of perturbed Hamiltonian systems. Note that the
Landau-Lifshitz equation is a Hamiltonian system, with canonically
conjugate variables $\cos(\Th)$, $\ph$. That is, the Landau-Lifshitz
system may be written as $\frac{ \partial \cos\Th} {\partial t} =
\frac{ \delta \mathcal{E}} {\delta \ph}$ and $\frac{ \partial \ph}
{\partial t} =- \frac{ \delta \mathcal{E}} {\delta \cos \Th}$, where
the right hand sides are expressed in terms of variational derivatives
of the energy $\mathcal{E}$, defined later in eq.~\eqref{eq:energy_def}. In
such systems, it is more convenient to execute perturbation theory in
the Hamiltonian variables.  It is also possible to write down
modulation equations for solitons with parameters related in a
particular way. This statement will be made more precise later in this
section.

The basic procedure is to allow the parameters to vary on a time scale
proportional to the strength of the perturbation, $\epsilon$. By
allowing the parameters to vary in this way, additional degrees of
freedom are introduced which can be used to resolve the difficulties
arising from singular perturbations. Expanding about the soliton
solution in an asymptotic series, one obtains a linear problem at
order $\epsilon$. In general, this linear equation will not admit
solutions bounded in time. However, as utilized in
\cite{weinstein1985modulational}, a solvability condition exists
whereby bounded solutions are assured, guaranteeing that the linear
problem at order $\epsilon$ does not break the asymptotic
ordering. Imposing these conditions leads to the modulation
equations. This procedure is equivalent to projecting the solution of
the perturbed model onto the family of solitons. While one might wish
to then solve the linear equation at order $\epsilon$ to obtain a
further correction, we will not do that in this work. As will be
demonstrated by the examples in Sec.~\ref{sec:applications}, quite
satisfactory predictions can be made considering only the leading
order dynamics.

A Hamiltonian system requires a real inner product space, $X$; a
nonlinear functional, $H:X\rightarrow \mathbb{R}$; and a skew adjoint
operator $J:X \rightarrow X$. We will use the notation $\langle \cdot
, \cdot \rangle$ for the inner product on $X$.  The standard form for
a Hamiltonian system is
\begin{equation}
\frac{ \partial z}{\partial t} = J \grad H(z)
\label{eq:classic_hamiltonian}
\end{equation}
where $z\in X$ is referred to as the state variable. We refer to $H$
as the Hamiltonian, which is often assigned the physical meaning of
energy since it is automatically a conserved quantity of such a
system. In this context, by $\grad H$ we mean the first variation of
this nonlinear functional and by $\Delta H$ we mean the second variation
(both taken with respect to the state variable, $z$). We consider here
Hamiltonians which depend explicitly upon additional parameters,
$\mathbf{q} \in \mathbb{R}^m$ ($m$ is the number of such
parameters). Such parameters may arise due to a change of coordinates,
such as to a comoving reference frame.  For the examples which arise
in this work, the parameters $\mathbf{q}$ arise from just such a
transformation, so we will typically refer to these parameters as
``frequencies".

We assume here that \eqref{eq:classic_hamiltonian} admits a solitary
wave solution, $u$. 
\begin{equation}
  \label{eq:zeroth_order}
  0  = J \grad H(u,\mathbf{q}).
\end{equation}
If $H$ depends on $\mathbf{q}$, naturally $u$ will depend on
$\mathbf{q}$ as well. Typically, the parameters $\mathbf{q}$ do not
provide a full parameterization of the solitary wave manifold due to
underlying symmetries in the equation such as translation
invariance. Accordingly, we will allow $u$ to depend on a separate set
of parameters $\mathbf{r}\in \mathbb{R}^s$ ($s$ is the number of such
parameters). For reasons that will become clear in later examples we
refer to these parameters as ``phases''.  Therefore, the solitary wave
can be written $u = u(\mathbf{x}; \mathbf{q},\mathbf{r})$.  Often
times, the Hamiltonian system \eqref{eq:classic_hamiltonian} admitting
solitary wave solutions \eqref{eq:zeroth_order} is idealized,
neglecting important physical effects. While some perturbations may
give rise to a different Hamiltonian system, in general such effects
do not need to preserve the Hamiltonian structure. We will treat both
cases the same by introducing a small perturbation into the equation
itself.  The perturbed model is
\begin{equation}
\frac{ \partial z}{\partial t} = J \grad H(z,\mathbf{q}) + \epsilon P
\label{eq:perturbed_hamiltonian}
\end{equation}
where $0<\epsilon\ll 1$ and $P$ is a perturbation. The parameters
$\mathbf{q},\mathbf{r}$ are allowed to vary on a slow time scale, $T =
\epsilon t$. We restrict to perturbations which depend explicitly on
time only through this slow time variable, $T$.  In this case,
ordinary differential equations governing the evolution of these
parameters can be determined according to the following theorem.
\begin{theorem}
\label{thm:main_result}
Given the perturbed Hamiltonian system \eqref{eq:perturbed_hamiltonian}.   If
\begin{enumerate}
\item The solitary wave solution, $u$, exists for the unperturbed
  system \eqref{eq:perturbed_hamiltonian}, $\epsilon = 0$, and is independent of $t$.
\item $J$ is invertible.
\item $\Delta H\bigg|_{z = u}$ is self-adjoint for all admissible $\mathbf{q}$.
\item  $\forall\;1\le k \le m$, $\exists\; 1\le j \le s$ such that
  $\frac{\partial}{ \partial q_k}  \grad H(z,\mathbf{q})\bigg|_{z = u}
  = - J^{-1} \frac{\partial u}{ \partial r_j} $ 
\end{enumerate}
then letting $\mathbf{v} = \left[\mathbf{r}, \mathbf{q} \right] ^T \in
\mathbb{R}^{s+m}$, 
the modulation equations are
\begin{equation}
  \begin{array}{rcl}
    \ds \left(\sum_{i = 1}^{s + m }  \left\langle J^{-1}
        \frac{\partial u }{\partial v_i}, \frac{\partial u }{\partial
          v_j} \right\rangle \frac{d v_i}{d T}  \right) %
    &\ds= &\ds\left\langle J^{-1} P, \frac{\partial u}{ \partial v_j} 
    \right \rangle\end{array} 
  \label{eq:hamiltonian_modulations}
\end{equation}
\end{theorem}

Equation \ref{eq:hamiltonian_modulations} is consistent with previous
general results when applied to Hamiltonian systems
\cite{keener_solitons_1977}. The assumptions of
Theorem~\ref{thm:main_result} may seem restrictive at first, but these
conditions are frequently met in physical systems of interest. In all
systems under consideration here there does exist a solitary wave
solution. These solutions generically depend on time, but for the case
of a single solitary wave solution, transforming to the reference
frame moving, rotating, and/or precessing with the solitary wave can
eliminate this explicit dependence on time. Such a transformation will
introduce parameters in $\mathbf{q}$ and alter the Hamiltonian but
leaves the Hamiltonian structure intact.

The second does offer a restriction. For instance, in the
Korteweg-de-Vries equation, $J$ does not admit a bounded inverse and
correspondingly the modulation equations require additional
considerations \cite{ablowitz1981solitons}. Nevertheless, formal
calculations are possible and $J$ is frequently invertible for
Hamiltonian systems (as it is, e.g., for NLS and the Landua-Lifshitz
equation).

With appropriate restrictions on the Hamiltonian, the third assumption
always holds. The self-adjoint property of the second variation
essentially follows from the same calculation which proves the
equality of mixed partial derivatives in finite-dimensional
calculus. More care needs to be taken in the corresponding calculation
on function spaces, but the Hamiltonians derived in physically
relevant systems typically are well enough behaved.

The fourth assumption is restrictive and may seem obscure. However,
the parameters of the soliton are often speeds or frequencies. These
parameters are typically linked to initial positions or initial phase
values so that $\mathbf{q}$ and $\mathbf{r}$ have the same length ($s
= m$). In such cases, the dependence of the soliton on the parameters
in the laboratory frame will be in the form $\mathbf{r} + t
\mathbf{q}$. From this temporal dependence, the relations in
Assumption (iv) follow directly.

\subsection{Derivation of Equation \eqref{eq:hamiltonian_modulations}}
Theorem \ref{thm:main_result} relies on the following lemma. 
\begin{lemma}
  \label{lem:solvability}
Let $X$ be a Hilbert space. Let $A$ be a linear operator mapping $X$ to itself. Let $f \in X$.  Let $A^\dagger$ be the adjoint of $A$, i.e. the unique linear operator satisfying $\langle A^\dagger x, y\rangle = \langle x, A y\rangle$ for all $x,y\in X$. Define $\vp:[0,\infty)\rightarrow X$ as the solution of the initial value problem
  \begin{equation}\begin{cases} \pd{\vp}{t} = A \vp + f \\
      \vp(0) = \vp_0 \in X .
    \end{cases}\label{eq:solvability_equation}
  \end{equation}
  Let $\mu_{-1} =0$ and $A^\dagger \mu_{i} = \mu_{i-1}$ for $0\le i
  \le N$, where $N$ denotes the highest integer such that
  $(A^{\dagger})^N$ has nontrivial kernel. Then $\vp(t)$ will not be
  bounded in time unless $ \inprod{\mu_{i-1},\vp_0} + \inprod{\mu_{i},
    f} = 0$ for $0\le i \le N$.
\end{lemma} 

This lemma is a minor generalization of the solvability condition
proven in \cite{weinstein1985modulational}. There are a few key
limitations which may not be clear upon first reading the statement of
the lemma itself. First, $A$ and $f$ are assumed to be independent of
time $t$. Second, all assumptions of smoothness of $f$ are
bound up in the choice of $X$ which is problem specific. In the
context of Hamiltonian systems, $X$ is given and the required
smoothness of $f$ is clear. In our intended application,
Eq. \eqref{eq:solvability_equation} arises from a linearization of a
nonlinear problem about a given state. In this case, $A$ and $f$ are
given, but not $X$. In order that Lemma \ref{lem:solvability} applies,
there must exist an $X$ which makes $A$ and $f$ compatible, and it
will be in that sense which $\vp(t)$ remains bounded in time. 
From here on out, we assume sufficient smoothness in our perturbation
such that a Hilbert space is naturally chosen. For the perturbations
we investigate in Section \ref{sec:applications}, this is the case.
Finally, the details of defining the adjoint of the unbounded operator
$A$ are not considered here but can be handled in a standard manner;
see, e.g., \cite{weinstein1985modulational}.
  
The proof of Theorem ~\ref{thm:main_result} proceeds by substituting the
ansatz
\begin{equation}
  \label{eq:ansatz}
  z= u(\mathbf{r}(T), \mathbf{q}(T)) + \epsilon u_1(\mathbf{x},t,T) +
  \bo{\epsilon^2} 
\end{equation}
into \eqref{eq:perturbed_hamiltonian}. Expanding in powers of
$\epsilon$, the first order equation becomes
\begin{equation}
\label{eq:leading_order}
\bo{\epsilon}: \hspace{5mm} \frac{\partial u_1}{\partial t} =  J
\Delta H(u,\mathbf{q}) u_1 - \frac{\partial u}{\partial \mathbf{r}} \frac{d
  \mathbf{r}}{d T}  - \frac{\partial u}{\partial \mathbf{q}} \frac{d
  \mathbf{q}}{d T}  + P 
\end{equation}
Note that Eq. \eqref{eq:leading_order} is of the form in Lemma
\ref{lem:solvability} ($A= J \Delta H(u,\mathbf{q})$, $f = P- \frac{\partial
  u}{\partial \mathbf{r}} \frac{d \mathbf{r}}{d T} - \frac{\partial
  u}{\partial \mathbf{q}} \frac{d \mathbf{q}}{d T}$). In order that
the expansion in \eqref{eq:ansatz} remain asymptotically ordered, it
is necessary that $u_1(x,t,T)$ remain $\bo{1}$ for sufficiently long
times. Lemma \ref{lem:solvability} thus gives a condition that must be
satisfied. It remains to characterize the generalized nullspace of $(J
\Delta H(u,\mathbf{q}))^\dagger$. Note that since $\Delta
H(u,\mathbf{q})$ is self-adjoint, $(J \Delta H(u,\mathbf{q}))^\dagger
= -\Delta H(u,\mathbf{q}) J$.

Differentiating \eqref{eq:zeroth_order} with respect to the parameter
$r_j$ for $1\le j \le s$ and applying $J^{-1}$ to the result yields
$\Delta H(u,\mathbf{q}) \frac{\partial u } {\partial r_j} = 0$. It
follows that $ J^{-1} \frac{\partial u } {\partial r_j} $ is in the
kernel of $(J \Delta H(u,\mathbf{q}))^\dagger $ for all $j$.
Differentiating \eqref{eq:zeroth_order} with respect to the parameter
$q_k$ for $1 \le k \le m$ yields
\begin{equation}
  \label{eq:3}
  \Delta H(u,\mathbf{q}) \frac{\partial u }
  {\partial q_k} + \frac{\partial}{\partial q_k} \grad
  H(z,\mathbf{q})\bigg|_{z = u} = 0 .
\end{equation}
Utilizing assumption (iv), we can replace the second term in
\eqref{eq:3} so that there is some $j$ with
 \begin{equation}
   \label{eq:generalized_eigenvector}
   \Delta H(u,\mathbf{q}) J  \left(J^{-1} \frac{\partial u } {\partial
       q_k} \right) = J^{-1} \frac{\partial u } {\partial r_j}. 
 \end{equation}
 Hence, $J^{-1} \frac{\partial u } {\partial q_k} \in \ker (\Delta
 H(u,\mathbf{q}) J )^2$ and therefore in the generalized
 nullspace. These two sets of vectors do not necessarily characterize
 the full generalized nullspace; however, these offer a
 sufficient number of constraints to uniquely determine the modulation
 system. Requiring that $f = P- \frac{\partial u}{\partial \mathbf{r}}
 \frac{d \mathbf{r}}{d T} - \frac{\partial u}{\partial \mathbf{q}}
 \frac{d \mathbf{q}}{d T}$ be orthogonal to $J^{-1} \frac{\partial u }
 {\partial q_k}$ and $J^{-1} \frac{\partial u } {\partial r_j}$ yields
 equations \eqref{eq:hamiltonian_modulations}.  The modes $J^{-1}
 \frac{\partial u } {\partial q_k}$ and $J^{-1} \frac{\partial u }
 {\partial r_j}$ may not give rise to a complete characterization of
 the nullspace. As a result, Eqs. \eqref{eq:hamiltonian_modulations}
 are only a necessary but not sufficient condition to prevent
 secular growth.

\section{Approximate Propagating Droplet}
\label{sec:droplet}

Now that we have the general formulation of the soliton modulation
equations for perturbed Hamiltonian systems in
eq.~\eqref{eq:hamiltonian_modulations}, we would like to apply them to
the magnetic droplet soliton solution of
eq.~\eqref{eq:nondimensional_ll}.  For this, we will need to compute
derivatives of the droplet with respect to its parameters as well as
associated inner products.  This could be performed numerically with a
``database'' of droplet solutions as in \cite{hoefer2012propagation}.
Here, we obtain an explicit, analytical formulation of the modulation
equations in the strongly nonlinear, moving droplet regime.  But before we can
determine the modulation equations, we need an explicit representation
of the propagating droplet itself.  In this section, we derive an
approximate solution to eq.~\eqref{eq:nondimensional_ll} when
$\epsilon = 0$, a restriction we maintain for the remainder of this
section.  The solution describes a slowly moving droplet with
frequency just above the Zeeman frequency.  A droplet solution can be
characterized by six parameters: its precession frequency $\omega$
above the Zeeman frequency $h_0$ in these non-dimensional units,
propagation velocity $\mathbf{V}=[V_x,V_y]^T$, initial phase $\Phi_0$,
and the coordinates of the droplet center $\boldsymbol{\xi} =
[\xi_x,\xi_y]^T = \mathbf{V} t + \mathbf{x}_0$.

Approximate droplet solutions have been found in two regimes: (i) $0 <
1 - \omega - |\mathbf{V}|^2/4 \ll 1$, near the linear (spin-wave) band
edge corresponding to propagating, weakly nonlinear droplets
approximated by the NLS Townes soliton \cite{ivanov2001small} (ii) $0
< \omega \ll 1$ with zero velocity corresponding to stationary,
strongly nonlinear droplets approximated by a circular domain wall
\cite{kosevich__1986,ivanov1989two}.  We will focus here on large
amplitude propagating solitons where the magnetization is nearly
reversed because experiments operate in this regime.  Note, however,
that the weakly nonlinear regime could also be studied.  The defining
equation for the droplet can be formulated as a boundary value problem
by expressing the magnetization in spherical variables $\mathbf{m} =
[\sin(\Th) \cos(\ph), \sin(\Th) \sin(\ph), \cos(\Th)]$ in the frame
moving and precessing with the soliton $\Th \to \Th(\mathbf{x} -
\boldsymbol{\xi})$, $\ph \to \ph_0 + (h_0 + \omega)t + \ph(\mathbf{x}
- \boldsymbol{\xi})$:
\begin{equation}
\ds
\left\{\begin{array}{rl}
    - \sin(\Th)\mathbf{V} \cdot \grad \Th &= \divg\left( \sin^2\Th
      \grad \ph) \right)\\[3mm] 
    \sin(\Th) ( \w - \mathbf{V} \cdot \grad \ph) &=  -\lap \Th +
    \frac{1}{2} \sin(2 \Th) (  1 + \abs{\grad \ph}^2)\\[3mm] 
    \underset{\abs{\mathbf{x}} \rightarrow \infty}{\lim} \grad \ph
    & = -\frac{\mathbf{V}}{2}, \quad \underset{\abs{\mathbf{x}}
      \rightarrow \infty}{\lim} \Th
    = 0 .
  \end{array} \right.
\label{eq:droplet_pde}
\end{equation}

This problem can be further simplified by exploiting the invariance of
Eq.~\eqref{eq:droplet_pde} under rotation of the domain to align the
$x$-axis with the propagation direction. In this coordinate system,
$\mathbf{V} = V \mathbf{\hat{x}}$.  Adding the assumptions of small
frequency and propagation speed, a simple correction to the known,
approximate stationary droplet can be found (see Appendix
~\ref{app:approximate_droplet}).

\begin{align}
  \Th = \cos^{-1}\left(\tanh\left(\rho - \frac{1}{\w} \right) \right)
  + \bo{\w^2, V^2} \label{eq:approxtheta}\\ 
  \ph = \ph_0 + (h_0 + \w) t - \frac{ V }{\w^2 } \cos(\vph) + \bo{
    \frac{V}{\w}}\label{eq:approxphi}.
\end{align}
Above, $(\rho, \vph)$ are polar variables for the plane, whose origin
is centered on the droplet. That is $\rho = \sqrt{(x -\xi_{x})^2 + (y
  - \xi_{y})^2} $ and $\vph = \arctan\left( \frac{y - \xi_{y}}{x -
    \xi_{x}}\right)$.  See Fig.~\ref{fig:droplet} for a visualization
of this approximate solution.  This approximation is valid so long as
\begin{equation}
  \label{eq:2}
  0 \le \abs{V} \ll \w, \quad 0 < \omega \ll 1 .
\end{equation}
As for the stationary case, the propagating droplet can be viewed as a
precessing, circular domain wall with a radius that is the inverse of
the frequency.  The new term $-V\cos (\varphi) /\omega^2$ reveals the
deviation of the propagating droplet's phase from spatial uniformity.
While the relations in \eqref{eq:2} may, at first, seem overly
restrictive, we will show that important and practical information
about propagating droplets can be obtained in this regime.  This
approximate solution offers both an error estimate and is amenable to
further analysis in the context of the perturbed Landau-Lifshitz
equation \eqref{eq:nondimensional_ll}.  Furthermore, it provides a
significant improvement over the approximate droplets used in past
numerical experiments \cite{piette1998localized} when the asymptotic
relations \eqref{eq:2} hold.

Another important property of Eq.~\eqref{eq:nondimensional_ll} in the
$\epsilon=0$ case is that it admits conserved quantities, including 
the total spin
\begin{equation}
  \mathcal{N} = \int_{\mathbb{R}^2} (1 - m_z)d \mathbf{x} ,
\label{eq:total_spin_def}
\end{equation}
the momentum
\begin{equation}
  \boldsymbol{\mathcal{P}} = \int_{\mathbb{R}^2}\left(  \frac{m_y \grad m_x - m_x
      \grad m_y}{1+ m_z}\right)d \mathbf{x} ,
 \label{eq:momentum_def}
\end{equation}
and the total energy,
\begin{equation}
  \mathcal{E} =  \frac{1}{2} \int_{\mathbb{R}^2} \left(
    \abs{\grad{\mathbf{m}}}^2 + ( 1 - m_z^2) \right) d \mathbf{x} +
  \frac{1}{2} h_0 \mathcal{N} .
\label{eq:energy_def}
\end{equation}
These quantities are not independent for the droplet itself, since the
droplet is the energy minimizing solution constrained by the total
spin and momentum.  Utilizing the approximate form
\eqref{eq:approxtheta}, \eqref{eq:approxphi} for the droplet, a map
can be constructed between its parameters and the conserved
quantities.  Evaluating the integrals in
Eqs.~\eqref{eq:total_spin_def}-\eqref{eq:energy_def} at the
approximate droplet, we obtain
\begin{align} 
  \mathcal{N} &= \frac{2 \pi}{\w^2},  \label{eq:total_spin_approx}\\
  \boldsymbol{\mathcal{P}} &= \frac{2 \pi}{\w^3}
  \mathbf{V} \label{eq:momentum_approx} \\ 
  \mathcal{E} & = \frac{\pi}{\w^3}\left(  \abs{ \mathbf{V}}^2 + 4 \w^2
  + h_0 \omega \right) \label{eq:energy_approx}. 
\end{align}
where higher order terms in $\w$ and $\abs{\mathbf{V}}$ have been
neglected.  These formulae extend the predictions for stationary
droplets, see, e.g., \cite{kosevich1990magnetic}, and offer an analogy
to classical particle dynamics. Rewriting $\mathcal{E}$ in terms of
the other conserved quantities, we obtain
\begin{equation}
  \mathcal{E}=  \sqrt{2 \pi} \left( \frac{1}{2}
    \frac{\abs{\boldsymbol{\mathcal{P}}}^2}{ \mathcal{N}^{
        \frac{3}{2}}} + 
    \mathcal{N}^{\frac{1}{2}}
  \right) + \frac{1}{2} h_0 \mathcal{N} .\label{eq:alt_energy_rep}
\end{equation}

By analogy to classical systems, we can interpret $ \sqrt{2 \pi}
\abs{\boldsymbol{\mathcal{P}}}^2/2\mathcal{N}^{ \frac{3}{2}}$ as the
kinetic energy of the droplet, $\sqrt{2 \pi}
\mathcal{N}^{\frac{1}{2}}$ as the droplet's potential energy due to
precession, and $h_0 \mathcal{N}/2$ as the Zeeman energy of the
droplet with the net dipole moment $\mathcal{N}$. Inspection of the
kinetic energy term shows that
\begin{equation}
  \label{eq:1}
  m_{\rm eff} = \frac{\mathcal{N}^{3/2}}{\sqrt{2 \pi}} =
  \frac{2\pi}{\omega^3} 
\end{equation}
serves as the effective mass for the droplet. Therefore, the $0 <
\omega \ll 1$ regime corresponds to droplets with large mass.  This is
a natural interpretation since it is the precession of the droplet
which determines its size and prevents the structure from collapsing
in on itself. On the other hand, eq.~\eqref{eq:momentum_approx}
implies that the slowly propagating $|\mathbf{V}| \ll \omega$ regime
supports droplets with up to $|\boldsymbol{\mathcal{P}}| =
\mathcal{O}\left(\frac{1}{\omega}\right)$ momenta.  We will return to this observation of an
effective mass for the droplet in Section
\ref{sec:applications}\ref{sec:applied_field}, where we consider the
dynamical equations induced by spatial inhomogeneity in the external
magnetic field.

One description of the magnetic droplet is as a bound state of magnons
\cite{kosevich1990magnetic}. It is then natural to interpret the
potential energy $\sqrt{2 \pi} \mathcal{N}^{\frac{1}{2}}$ as the
energy released by decay into these constituent ``subatomic
particles''.  The expressions \eqref{eq:total_spin_approx} and
\eqref{eq:momentum_approx} can also be utilized to verify the
Vakhitov-Kolokolov soliton stability criteria
\cite{vakhitov_stationary_1973,grillakis1990stability} for a
propagating droplet (see \cite{hoefer2012propagating}), namely that
$\mathcal{N}_\omega = -4\pi/\omega^3 < 0$ and $\mathcal{N}_\omega
\nabla_V \cdot \boldsymbol{\mathcal{P}} - \nabla_V \mathcal{N} \cdot
\boldsymbol{\mathcal{P}}_\omega = -8\pi^2/\omega^6 < 0$, as required.

For the remainder of this work, we will use the approximate droplet in
eqs.~\eqref{eq:approxtheta}, \eqref{eq:approxphi}.

\section{General Modulation Equations for Propagating Droplets}
\label{sec:general_result}
With the results of the previous sections, we now have developed
sufficient tools to derive the droplet modulation equations. Previous
attempts to do this have been limited either to a partial set of
equations for $\mathbf{V}$ and $\omega$ only, computed numerically
\cite{hoefer2012propagation}, or stationary droplet equations
\cite{bookman2013analytical}.  The results of Sections
\ref{sec:hamiltonian_systems} and \ref{sec:droplet} enable us to
determine the slow evolution of all six soliton parameters due to the
perturbation $\mathbf{p}$ in eq.~\eqref{eq:nondimensional_ll}.  The
calculation, not presented, requires some care in preserving the
appropriate asymptotic relations \eqref{eq:2}.  From expressions
\eqref{eq:hamiltonian_modulations}, \eqref{eq:approxtheta}, and
\eqref{eq:approxphi}, we obtain the droplet soliton modulation
equations
\begin{align}
\dot{\ph}_0 & =\frac{1}{4 \pi } \intrr{(\mathbf{V} \cdot
  \hat{\boldsymbol{\rho}}) \sinth p_\Theta} 
+\frac{\w}{4 \pi } \intrr{  \sech\left(\rho - \frac{1}{\w} \right)
  p_{\cph}}, 
\label{eq:gen_ph_eqn}\\[2mm]
\dot{\boldsymbol{\xi}} & =\frac{\mathbf{V}}{\epsilon} +
\frac{\w}{2 \pi }\intrr{ \sinth \hat{\boldsymbol{\rho}} \: p_\Theta  } ,
\label{eq:gen_x0_eqn}\\[2mm]
\dot{\w} & = -\frac{ \omega ^3}{4 \pi } \intrr{ \sinth p_\Th } ,
\label{eq:gen_w_eqn}\\[2mm]
\dot{\mathbf{V}} & =-\frac{\omega ^2}{2   \pi } \intrr{
  \left(\frac{3}{2} \mathbf{V} - \frac{ \left( \mathbf{V} \cdot
        \hat{\boldsymbol{\vph}} \right) }{\rho
      \omega}\hat{\boldsymbol{\vph}} \right)  \sinth p_\Th}
- \frac{\omega ^3}{2 \pi }\intrr{ \sech\left(\rho - \frac{1}{\w}
  \right) \hat{\boldsymbol{\rho}} \:p_{\cph}} ,
\label{eq:gen_v_eqn}
\end{align}
where the over dot denotes differentiation with respect to $T$.  This
general set of equations is the main result of this work.  They are
asymptotically valid when
\begin{equation}
  \label{eq:6}
  0 < \e \ll 1, \quad T \ll \epsilon^{-1}, \quad 0 \le
  |\mathbf{V}|\ll \omega \ll 1 .
\end{equation}
The perturbation components $p_\Th$, $p_{\cph}$ are to be evaluated
with the approximate droplet solution \eqref{eq:approxtheta},
\eqref{eq:approxphi}.  Some explanation of the perturbation components
and the unit vectors $\hat{\boldsymbol{\rho}}$,
$\hat{\boldsymbol{\varphi}}$ is warranted.  The magnetization
$\mathbf{m}$ has the unit sphere $\mathbb{S}^2$ as its range.  We can
therefore define the standard, right-handed, orthonormal, spherical
basis $\left \{ \hat{\mathbf{r}}, \hat{\boldsymbol{\cph}},
  \hat{\boldsymbol{\Th}}\right \}$ for $\mathbb{R}^3$ where
$\hat{\mathbf{r}} = \mathbf{m}$ is the radial unit vector,
$\hat{\boldsymbol{\cph}}$ is the azimuthal unit vector, and
$\hat{\boldsymbol{\Th}}$ is the polar unit vector.  The components of
$\mathbf{p}$ in this ``magnetization centered'' basis are
\begin{equation}
  \label{eq:4}
  0 = \hat{\mathbf{r}} \cdot \mathbf{p}, \quad p_\cph = \hat{\boldsymbol{\cph}}
  \cdot \mathbf{p}, \quad p_\Th = \hat{\boldsymbol{\Th}} \cdot \mathbf{p} .
\end{equation}
On the other hand, the domain $\mathbb{R}^2$ has the standard
orthonormal, polar basis $\left
  \{\hat{\boldsymbol{\rho}},\hat{\boldsymbol{\varphi}}\right \}$ where
$\hat{\boldsymbol{\rho}}$ and $\hat{\boldsymbol{\varphi}}$ are the
radial and azimuthal unit vectors, respectively.  This corresponds to
the ``domain centered'' basis.  It is important not to confuse the
domain $\mathbb{R}^2$ and range $\mathbb{S}^2$ of $\mathbf{m}$.

In this general formulation, we have neglected spatial inhomogeneity
of the perpendicular magnetic field magnitude $h_0$.  In Section
\ref{sec:applications}~\ref{sec:applied_field}, we will incorporate
inhomogeneity as a perturbation with nonzero $p_\cph$ component.  Even
with an inhomogeneous magnetic field, the total frequency of the
droplet $\Omega(T)$ is
\begin{equation}
  \label{eq:5}
  \Omega(T) = h_0(\epsilon \boldsymbol{\xi}(T),T) + \omega(T) + \e
  \dot{\cph}_0(T) .
\end{equation}
We see that variations in the initial phase $\cph_0$ provide a higher
order correction to the droplet frequency.  Additionally, the second
term on the right hand side of eq.~\eqref{eq:gen_x0_eqn} is a higher
order correction to the droplet's total velocity
$\dot{\boldsymbol{\xi}}$.  These higher order corrections have proven
to be of fundamental importance in the study of stationary droplets
\cite{bookman2013analytical} and beyond, see, e.g.,
\cite{ablowitz_perturbations_2011} for an application to NLS dark
solitons.

While quite general, these equations do not treat all
perturbations. It is important to note that the solvability condition
which gives rise to these equations applies for those perturbations
whose temporal dependence is on a slow time scale.  In the sections
that follow, we consider a range of physical perturbations that meet
this criterion in order to demonstrate the versatility of this
approach. However, some physical scenarios (such as an applied field
varying rapidly in time) might not satisfy this assumption. Such
perturbations may be regular perturbations and not induce dynamics
within the family of solitons so they will not be further discussed.

\section{Applications To Perturbed Systems}
\label{sec:applications}
In this section we analyze a range of perturbations to demonstrate the
versatility of this framework as well as to provide physical insights
into droplet dynamics.

\subsection{Slowly Varying Applied Field}
\label{sec:applied_field}

In practical applications, the magnetic field will typically have some
spatial variation whose scale is much larger than the scale of the
droplet, i.e., the exchange length divided by $\w$. For this, we
assume that $h_0 = h_0( \e \mathbf{x},\e t)$, $0< \epsilon/\omega \ll 1$.
This inhomogeneity is best treated by introducing an appropriate
perturbation $\mathbf{p}$ in eq.~\eqref{eq:nondimensional_ll}.
Expanding $h_0$ about the soliton center, $\bxi$,
\begin{equation}
  h_0(\e \mathbf{x},\e t) = h_0(\e \boldsymbol{\xi},\e t) + \e \tilde{\grad}
  h_0\big|_{\mathbf{x} = \bxi} \cdot ( \mathbf{x} - \bxi ) +
  \bo{\e^2},
\label{eq:expand_h0}
\end{equation}
where $\tilde{\grad}$ represents the gradient with respect to the slow
variable $\boldsymbol{X} = \e \boldsymbol{x}$.  Inserting the
expansion \eqref{eq:expand_h0} into the cross product $-\mathbf{m}
\times (h_0 \hat{\mathbf{z}})$ from eq.~\eqref{eq:nondimensional_ll}
introduces the perturbation
\begin{equation}
  \ds p_\Th = 0 \mbox{ and } p_\ph = \left(\tilde{\grad} h_0 \cdot
    \hat{\boldsymbol{\rho}} \right) \rho  .
  \label{eq:h0perturb}
\end{equation}
Substituting these into
eqs.~\eqref{eq:gen_ph_eqn}-\eqref{eq:gen_v_eqn} leads to Newton's
second law for the droplet center

\begin{equation}
  \frac{d^2\boldsymbol{\xi}}{d t^2} = \e \frac{d
    \mathbf{V}}{d T} = -\w \grad h_0 ,
\label{eq:grad_h0_eqn}
\end{equation}
Note that $\grad$ here represents the gradient with respect to the
fast variable $\mathbf{x}$, distinguishing it from $\tilde{\grad}$.
The phase $\ph_0$ and frequency $\omega$ are unchanged by the field
gradient.

\begin{figure}[htbp]
  \centering
  \includegraphics{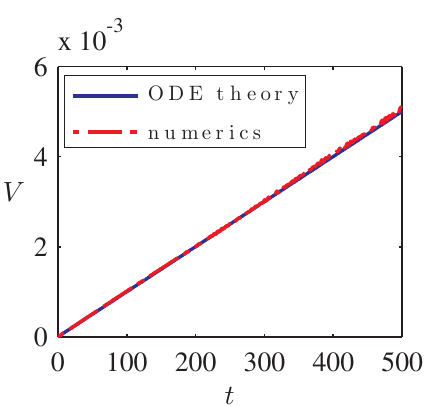}
  \caption{Acceleration of the droplet due to the inhomogeneous
    magnetic field $h_0 = 0.5-10^{-4}x$ with $\omega(0) = 0.1$ and
    $\abs{\mathbf{V}(0)} = 0$.  The exact solution to
    eq.~\eqref{eq:grad_h0_eqn} (solid) compares favorably to direct
    numerical simulations of the PDE (dashed). }
  \label{fig:field_gradient_plot}
\end{figure}

A favorable comparison of direct numerical simulations for
eq.~\eqref{eq:nondimensional_ll} (see Appendix \ref{app:mu-magnetics})
with the solution to \eqref{eq:grad_h0_eqn} is shown in
Fig.~\ref{fig:field_gradient_plot}.
We now demonstrate that the explicit equation \eqref{eq:grad_h0_eqn}
agrees with the previous result in \cite{hoefer2012propagation}
obtained by perturbing conservation laws and integrating the equations
numerically.  Previously, the nontrivial dynamical equation was
$\frac{d \mathcal{P} }{ d t } = - \mathcal{N} \grad h_0$. We can
transform this equation into eq.~\eqref{eq:grad_h0_eqn} by using the
explicit formulae \eqref{eq:total_spin_approx},
\eqref{eq:momentum_def} for $\mathcal{N}$ and $\mathcal{P}$. Since
$\frac{d \w } {d T } = 0$ and $\mathcal{N}$ depends only on $\w$,
$\frac{d \mathcal{N} } {d T } = 0$.  Then
\begin{equation} 
  \frac{d \mathcal{\boldsymbol{P}} }{d T} =
  \frac{\mathcal{N}^{3/2}}{\sqrt{2\pi}} 
  \frac{d \mathbf{V}}{d T}  = \frac{m_{\rm eff}}{\e} \frac{d^2
    \boldsymbol{\xi}}{d t^2} = - \frac{\mathcal{N}}{\e} \grad h_0 . 
\end{equation} 
This is exactly \eqref{eq:grad_h0_eqn}.  The particle-like droplet
with mass $m_{\rm eff}$ in eq.~\eqref{eq:1} experiences a conservative
force due to the potential $\mathcal{N} h_0$.  This interpretation is
consistent with the analysis of the effective mass derived from the
kinetic energy in Section \ref{sec:droplet}.  Furthermore, it
demonstrates that a droplet in a magnetic field gradient behaves
effectively like a single magnetic dipole with net dipole moment
$\mathcal{N}$.  

The effect of an inhomogeneous magnetic field on a massive
two-dimensional droplet is markedly different from its effect on a
one-dimensional droplet \cite{kosevich1998magnetic} and a vortex
\cite{papanicolaou1991dynamics}.  A one-dimensional droplet
experiences periodic, Bloch-type oscillations for a magnetic field
with constant gradient, while a magnetic vortex exhibits motion
perpendicular to the field gradient direction.

\subsection{Damping}
\label{sec:damping}
In \cite{hoefer2012propagation}, it was observed that the droplet
accelerates as it collapses in the presence of damping alone. The
framework presented here offers an analytical tool to understand this
slightly counterintuitive result, namely that damping can cause the
otherwise steady droplet to speed up. The relevant contributions to
eq.~\ref{eq:nondimensional_ll} are
\begin{equation}
  \ds p_\Th = - ( \w +h_0 - \mathbf{V} \cdot \grad \ph)
  \sin(\Theta)\mbox{ and } p_\ph = - \mathbf{V} \cdot \grad \Th 
  \label{eq:damping_perturbation} 
\end{equation}
where the small parameter $\e$ is the Landau-Lifshitz magnetic damping
parameter, usually denoted $\alpha$.  In many practical situations,
the damping parameter is quite small.

Evaluation of equations \eqref{eq:gen_ph_eqn}-\eqref{eq:gen_v_eqn} with
these perturbations yields two nontrivial equations
\begin{align}
  \frac{d \w}{d T} & = \omega^2 \left( \omega
    +h_0\right) \label{eq:damping_modw} \\[2mm]
  \frac{d \mathbf{V}}{d T} & = \omega \mathbf{V} \left( \omega
    + 2 h_0\right)\label{eq:damping_modv} .
\end{align}
These equations are again consistent with the numerical, perturbed
conservation law approach taken in \cite{hoefer2012propagation} when
evaluated at the approximate solution.

\begin{figure}[htbp]
  \centering
  \begin{tabular}{cc}
    \includegraphics{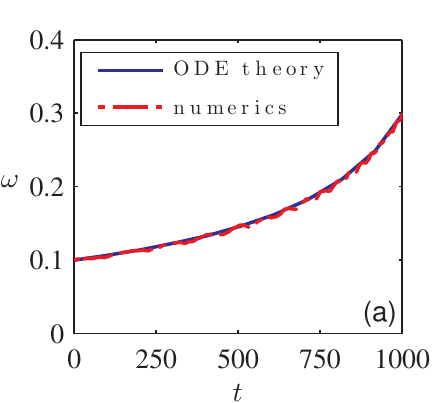}
    &\hspace{-3.5mm}  \includegraphics{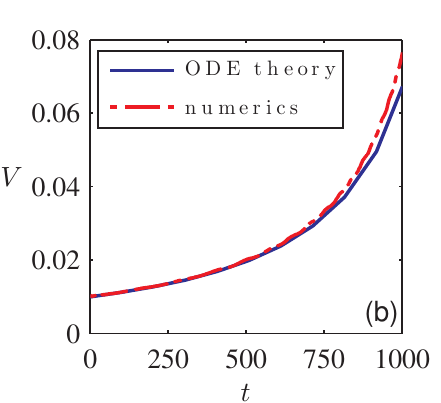} 
  \end{tabular}
  \caption{The evolution of droplet frequency (a) and velocity (b) due
    to damping for both numerical solutions of
    eqs.~\eqref{eq:damping_modw}, \eqref{eq:damping_modv} (solid) and
    direct numerical simulations of eq.~\eqref{eq:nondimensional_ll}
    (dashed) when $\epsilon = \alpha = 0.01$, $h_0 = 0.5$, $\omega(0)
    = 0.1$ and $\abs{\mathbf{V}(0)} = 0.01$.}
  \label{fig:damping_plots}
\end{figure}

We observe that the right hand sides of the modulation equations are
both positive for $h_0 > -\w/2$. Hence, the frequency and velocity
increase. Equation \eqref{eq:damping_modw} can be interpreted as a
dynamical equation for the droplet's mass $m_{\rm eff}$
(eq.~\eqref{eq:1}).  The mass is decreasing at a faster rate than the
velocity.  In light of the interpretation given in Section
\ref{sec:applications}\ref{sec:applied_field}, even though the droplet
is losing energy, it sheds mass fast enough that its acceleration is
not a contradiction.  In Fig. \ref{fig:damping_plots} we see good
agreement between the modulation theory and full micromagnetic
simulations.
 
Since Eq. \eqref{eq:damping_modw} decouples in this system, an
analytical solution can be found. Elementary application of partial
fractions yields an explicit solution in terms of the Lambert
W-function; however, the analysis is significantly simplified when
$h_0=0$. In this case, the analytical solution to
Eqs.~\eqref{eq:damping_modw}-\eqref{eq:damping_modv} is
\begin{align}
\omega(t) &= \frac{\w_0}{\sqrt{1-2  \alpha \w_0^2 t
  }},  \label{eq:damping_w_exact}\\ 
\mathbf{V}(t) &= \frac{\mathbf{V}_0}{\sqrt{1-2  \alpha \w_0^2 t
  }},   \label{eq:damping_V_exact} 
\end{align}
where $\w_0$ is the initial precession frequency and $\mathbf{V}_0$
the initial velocity. These expressions reveal two facts: a clear time
of breakdown for modulation theory and the existence of an adiabatic
invariant. Dividing Eq.~\eqref{eq:damping_w_exact} by the components
of Eq.~\eqref{eq:damping_V_exact} demonstrates that the quantities
$\w/V_x$ and $\w/V_y$ are constant in time.

\subsection{NC-STO and Spatially Inhomogeneous Applied Field}
\label{sec:NC_STO+Field}

So far, the examples we have chosen to focus on have not included
higher order contributions via the phase $\Phi_0$ and the second term
of the equation for the droplet center $\boldsymbol{\xi}$.  But many
perturbations and physical behaviors cannot be investigated without
these higher order terms. Consider the more complex system of a
nanocontact spin-torque oscillator (NC-STO), in which a polarized spin
current exerts a torque on the magnetization, the spin transfer torque
\cite{berger1996emission,slonczewski1996current}.  This forcing can be
confined to a localized region via a nanocontact
\cite{tsoi_excitation_1998,slonczewski_excitation_1999,rippard_direct-current_2004}.
Perturbations of this sort lead to dynamics within all the parameters
of the droplet.  In addition to spin torque, a droplet in a NC-STO
also experiences damping and it is precisely the balance between the
two that leads to the stable droplet observed in experiments.
Here, we consider the addition of weak spatial inhomogeneity of the
applied magnetic field. For simplicity we restrict our consideration
to a constant magnetic field gradient.

This investigation has broader implications for the practical use and
understanding of droplets in real devices.  We show in this section
that these three physical effects influence the system in competing
ways, which can balance, allowing for the existence of stable
droplets.  Alternatively, a strong enough field gradient can push the
droplet out of the NC-STO, giving rise to a previously unexplained
drift instability \cite{hoefer2010theory}. As seen in
Sec.~\ref{sec:applications}\ref{sec:damping}, damping decreases the
effective mass of the droplet. In
Sec.~\ref{sec:applications}\ref{sec:applied_field}, it was shown that
a field inhomogeneity accelerates the droplet while leaving the mass
of the droplet unaffected. The inclusion of forcing due to spin
transfer torque in a nanocontact opposes both of these effects. The
spin torque increases the droplet mass and generates an effective
restoring force that centers the droplet in the nanocontact region
\cite{bookman2013analytical}. Hence, there can exist a delicate
balance between all of these effects: the NC-STO restoring force
balancing the potential force due to the field gradient and the mass
loss due to damping balancing the mass gain due to
spin-torque. Previous studies have been unable to identify when such a
balance occurs and when it fails.  Here, we analytically demonstrate
stable droplets as fixed points of the modulation equations with all
of these perturbations.

Because the perturbation components $p_\Th$ and $p_\cph$ appear
linearly in the modulation equations
\eqref{eq:gen_ph_eqn}-\eqref{eq:gen_v_eqn}, we can simply add the
field inhomogeneity eq.~\eqref{eq:h0perturb} and damping
eq.~\eqref{eq:damping_perturbation} perturbations to those due to spin
torque \cite{stiles_spin_2006}.  Due to the presence of three
different perturbations, we no longer scale the perturbation
$\mathbf{p}$ in eq.~\eqref{eq:nondimensional_ll} by the single
parameter $\e$.  Rather, we set $\e = 1$ and introduce the small
parameters in $p_\Th$ and $p_\cph$ directly.  The perturbation
components are
\begin{align} 
  \label{eq:pth_ncsto}
  p_\Th &=   -\alpha \left( \w + h_0 - \mathbf{V} \cdot \grad
    \ph \right ) \sin 
  \Th + \sigma \mathscr{H}(\rho_* - r) \sin \Th,\\
  \label{eq:pph_ncsto}  
  p_\ph &= \left(\grad h_0 \cdot
    \hat{\boldsymbol{\rho}}\right)\rho - \alpha  \mathbf{V}
  \cdot \grad \Th  . 
\end{align}
The nanocontact where spin torque is active is assumed to be a circle
with radius $\rho_*$.  The coordinate $r$ in the argument of the
Heaviside function $\mathscr{H}$ is measured from the center of the
nanocontact, which differs from the coordinates $\rho$ and $\varphi$
which are measured from the center of the droplet. For simplicity, we
have neglected the spin torque asymmetry that introduces another
parameter into the analysis but does not appear to have a significant
effect on the dynamics \cite{hoefer2010theory}.  Experiments
\cite{mohseni2013spin,macia2014stable} and analysis
\cite{hoefer2010theory,bookman2013analytical} have shown that the
ratio of damping, $\alpha$, to forcing strength, $\sigma$
(proportional to current), is roughly order 1 for the existence of
droplets to be satisfied.  Thus $0 < \sigma \sim \alpha \ll 1$.  The
magnetic field is assumed to be linear
\begin{equation}
  \label{eq:7}
  h_0 = a + b x, \quad |b| \ll \w .
\end{equation}
We can restrict to droplet motion in the $\hat{\mathbf{x}}$ direction
only.  Insertion of the perturbations in Eqs.~\eqref{eq:pth_ncsto},
\eqref{eq:pph_ncsto} into the modulation equations
\eqref{eq:gen_ph_eqn}-~\eqref{eq:gen_v_eqn} results in the following
system
\begin{align}
  \ds \dot{\ph}_0 &= \ds \frac{\alpha b V}{2 \omega ^2}
  -\frac{\sigma V}{4 \pi}\intnco{\cos (\varphi )
    \text{sech}^2\left(\rho
      -\frac{1}{\omega }\right)}  \label{eq:dph0gen}\\[2mm]
  \ds \dot{\xi} & \ds = V-\frac{\alpha b }{\omega
  }+\frac{\sigma \omega}{2 \pi } \intnco{\cos (\varphi )
    \text{sech}^2\left(\rho -\frac{1}{\omega}\right)}
     \label{eq:dxigen}\\[2mm]
     \ds \dot{\w}&      \ds =
     \alpha  \omega^2  \left(   \omega+a \right)-\frac{\sigma
       \omega ^3}{4 \pi } \intnco{\text{sech}^2\left(\rho
           -\frac{1}{\omega }\right)} 
   \label{eq:dwgen} \\[2mm]
   \ds \dot{V}& \ds =
      - b\omega +\alpha V \omega
      \left(\omega+2 a   \right) 
    \ds 
    -\frac{ \sigma  V \omega }{4\pi
    }\intnco{\frac{ (3 \rho  \omega +\cos (2 \varphi )-1)}{ \rho
        } \text{sech}^2\left(\rho -\frac{1}{\omega
            }\right)} 
    \label{eq:dVgen}
\end{align}
where we set $\xi = \xi_x$, $V = V_x$, and the over dot denotes
differentiation with respect to $t$.  None of the right hand sides in
the equations above depend explicitly on the parameter $\Phi_0$ so
that the dynamics of the remaining parameters can be considered
separately. We ignore the evolution of $\dot{\ph}_0$ for the remainder
of the analysis noting that $\dot{\ph}_0$ corresponds to a small
frequency shift as in Eq.~\eqref{eq:5} that can be obtained from the
evolution of the other parameters by insertion into
Eq.~\eqref{eq:dph0gen}.

There is a complex interplay between the many small parameters in this
problem. Since we do not have access to an exact analytical solution,
it is necessary that these perturbations dominate over the error terms
in our approximate solution, while still remaining small. Since we
have $|V| \lesssim \w^2$ to keep an overall consistent error estimate
for the approximate droplet, we require that $\alpha, \sigma \ll
\w$. The variation in the applied field, $b$, is a more subtle and the
appropriate scaling will be determined by directly computing fixed
points.

The stationary droplet without a field gradient is stable when
centered on the nanocontact
\cite{hoefer2010theory,bookman2013analytical}.  This results from an
analysis of the stationary modulation equations which exhibit an
attractive, stable fixed point. Taking $\frac{d h_0}{d x} = b = 0$,
the modulation equations \eqref{eq:dxigen}-\eqref{eq:dVgen} for
propagating droplets exhibit the same fixed point $(\xi,\omega,V) =
(0,\omega_*,0)$ when damping balances forcing corresponding to the
current
\begin{equation}
\frac{\sigma }{\alpha }=\frac{2 \left(a+\omega _*\right)}{ 1+\omega _* \left( \log \left(\frac{1}{2} \text{sech}\left(\rho _*-\frac{1}{\omega _*}\right)\right)
+ \rho _* \tanh \left(\rho _*-\frac{1}{\omega
   _*}\right) \right)}.
   \label{eq:omega_fixed_pt}
\end{equation}
There is a saddle node bifurcation as $\sigma$ is increased, with
$\omega_* = \omega_*(\sigma)$ corresponding to the stable branch.  For
$\sigma$ sufficiently large, the stable branch quickly approaches
\begin{equation}
  \label{eq:6}
  \omega_* = \rho_*^{-1} + \mathrm{arctanh}\left ( \frac{2a
        \alpha}{\sigma}-1 \right )\rho_*^{-2} + \mathcal{O}\left (
    \rho_*^{-3} \right ), \quad \rho_* \gg 1, \quad 0 <
  \omega_* - \rho_*^{-1} \ll 1 .
\end{equation}
Near the critical value $\sigma = 2 a \alpha$, where the second term
is small, the asymptotic form is
\begin{equation}
  \label{eq:10}
  \omega_* = \rho_*^{-1} + \left ( \frac{2 a \alpha}{\sigma} -
    1 \right )\rho_*^{-2} + \left (\frac{2\alpha}{\sigma} + \ln 2
  \right ) \rho_*^{-3} +
  \mathcal{O} \left ( \rho_*^{-4} \right ), 
  \quad \left | \frac{2a \alpha}{\sigma} - 1 \right | = \mathcal{O}
  \left ( \rho_*^{-1} \right ) .
\end{equation}
Linearizing equations \eqref{eq:dxigen}-~\eqref{eq:dVgen} about this
fixed point, the Jacobian matrix is given by
\begin{align}
  \label{nc_sto_jac}
  J(0,\w_*, 0) &=  \left(
    \begin{array}{ccc}
      \lambda_1& 0 & 1
      \\
      0 &\lambda_2   & 0
      \\
      0 & 0 & \lambda_3 \\
    \end{array}
  \right), \\
  \label{eq:nc_sto_eig1}
  \lambda_1 &= -\frac{1}{2} \sigma \rho_* \omega_{*}
  \text{sech}^2\left(\rho_*-\frac{1}{\omega_{*}}\right) , \\
  \label{eq:nc_sto_eig2}
  \lambda_2 &= -\alpha  a \omega _*+\lambda _1+\frac{1}{2} \sigma  \omega _* \left(\tanh \left(\rho
   _*-\frac{1}{\omega _*}\right)+1\right),  \\
  \lambda_3 &= -2 \alpha \w_*^2  +\lambda_2 -\lambda_1.\label{eq:nc_sto_eig3} 
\end{align}
This linearization represents a generalization of that considered in
\cite{bookman2013analytical} where the motion was restricted to $V =
0$.  Since $\rho_* > \w_*^{-1}$, we observe that all eigenvalues are
negative when $\sigma > a \alpha$, so the fixed point is stable.  The
critical forcing value $\sigma = a \alpha$, below which the droplet
may be unstable could be considered as an estimate for the minimum
sustaining current of a droplet \cite{hoefer2010theory}.  Note,
however, that this is a dubious estimate due to $\omega_* -
\rho_*^{-1}$ not being a small quantity.  Utilizing the approximation
from Eq. \eqref{eq:10}, we find
\begin{equation}
  \label{eq:8}
  \begin{split}
    \lambda_1 &= -\frac{\sigma}{2} + \mathcal{O}\left (\rho_*^{-2} \right), \quad
    \lambda_2 \sim \lambda_1, \quad
    \lambda_3 =\left( -\alpha + \frac{\ln 2}{2} \sigma \right)\rho_*^{-2} +
    \mathcal{O}\left (\rho_*^{-4} \right) , 
  \quad \left | \frac{2a \alpha}{\sigma} - 1 \right | = \mathcal{O}
  \left ( \rho_*^{-1} \right ).
  \end{split}
\end{equation}
  
We now turn our attention to the case of a small field gradient $0 <
\abs{b} \ll 1$, where we observe the persistence of the droplet fixed
point for very small $\abs{b}$.  These fixed points exist as a balance
between the expulsive force provided by the field gradient and the
attractive force provided by the nanocontact. This attraction
manifests in the evolution of $\xi$ and so this balance can also be
viewed as a balance between leading order effects (in $V$) and higher
order effects (in $\xi$). Unlike the stationary fixed point, exact
analytical expressions for the fixed point cannot be found since the
droplet is no longer centered on the nanocontact ($\xi \not =
0$). Nevertheless, we can obtain the approximate form for these fixed
points as
follows.  
The structure of $J$ in Eq. \eqref{nc_sto_jac} yields very simple
predictions in the regime of small field gradient. The key observation
here is that the system of Eqs. \eqref{eq:dxigen}-\eqref{eq:dVgen} can be
written as
\begin{equation} 
  \label{eq:9}
  \begin{pmatrix} \dot{\xi} \\ \dot{\w} \\ \dot{V} \end{pmatrix} = F(\xi,\w,V)
  - b \begin{pmatrix}  \frac{\alpha}{ \w} \\ 0 \\ \w   \end{pmatrix}. 
\end{equation} 
By virtue of the stationary fixed point, $F$ satisfies $F(0, \w_*,0) =
0 $.  We now seek a fixed point that slightly deviates from the
stationary one according to
$\xi = b \xi_1 + \cdots $, $\w = \w_* + b \w_1 + \cdots $ and $V = b V_1
+ \cdots $.
\begin{figure}
  \hspace{-3mm}\begin{tabular}{ccc}
    \includegraphics{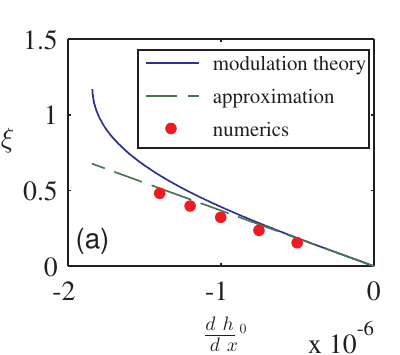}&\includegraphics{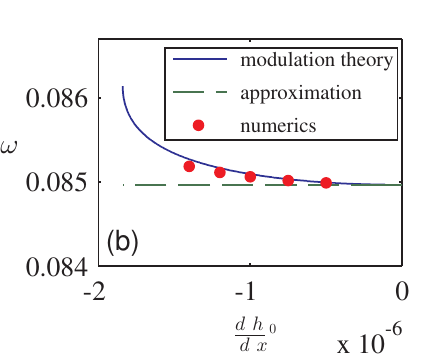} &
    \includegraphics{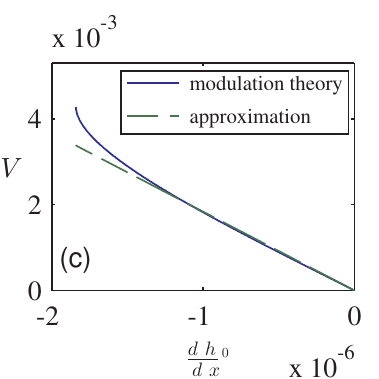} 
  \end{tabular}
  \caption{Fixed points from modulation theory, exact (solid) and
    approximate Eq.~\eqref{eq:14} (dashed), and direct numerical
    simulation of Eq.~\eqref{eq:nondimensional_ll} (circles) when
    $\alpha=\sigma=0.01$, $a = 0.5$, $\rho_* = 12$. In this case, the parameter $V$ cannot be extracted from direct numerical simulations without additional assumptions (see Appendix \ref{app:mu-magnetics}). Accordingly, this data is not presented in (c). }
   \label{fig:linear}
\end{figure}
Expanding and equating the right hand side of Eq.~\eqref{eq:9} to zero
gives 
the correction
\begin{equation}
  \label{eq:14}
  \begin{pmatrix} \ds \xi_1 \\ \ds \w_ 1\\ \ds
    V_1 \end{pmatrix} = J(0,\w_*, 0)^{-1}\begin{pmatrix} \frac{\alpha}{\omega_*} \\
    0\\ w_* \end{pmatrix}  = \begin{pmatrix} 
    \frac{\alpha}{\lambda_1 \w_*}-\frac{\w_*}{\lambda_1 \lambda_3}  \\ 0 \\ 
    \frac{\w_*}{ \lambda_3} \end{pmatrix} \sim
  \begin{pmatrix}
    \frac{4 \rho_*}{-2 \alpha\sigma + \sigma^2 \ln 2}  \\
    0 \\
   \frac{2 \rho _*}{ -2\alpha +\sigma \ln 2  }
  \end{pmatrix} 
  ,
\end{equation}
where the approximations \eqref{eq:10} and \eqref{eq:8} were used to
obtain the large $\rho_*$ estimate.  These approximations are valid so
long as $\sigma$ is more than $\rho_*^{-2}$ away from the critical
value $2\alpha/\ln 2 \approx 2.9 \alpha$.  Otherwise, higher order
terms in Eq.~\eqref{eq:8} would need to be considered.

As summarized in Fig.~\ref{fig:linear}, these simple expressions make
predictions in good agreement with the fixed points found by numerical
continuation in $b$ and those observed in long time micromagnetic
simulations of Eq.~\eqref{eq:nondimensional_ll} with perturbations
\eqref{eq:pth_ncsto} and \eqref{eq:pph_ncsto}.
The Jacobian matrix of Eqs. \eqref{eq:dxigen}-\eqref{eq:dVgen} can
also be numerically evaluated, showing that all eigenvalues are
negative, until continuation breaks down when one eigenvalue reaches
zero.  After this bifurcation, we do not find any fixed points. The
condition of this eigenvalue reaching zero then corresponds exactly to
the crossover where the attractive nanocontact is no longer strong
enough to balance the expulsive force supplied by the field gradient.
If we assume that a field gradient is strong enough to move the
droplet an order one distance, still small relative to the droplet
radius $\w_*^{-1}$, then we obtain a typical field gradient scaling $b
\approx \alpha \sigma/2\rho_*$.  This field gradient is very small.
For the example studied here, $b \approx 10^{-6}$ compared to the
NC-STO forcing magnitude $\sigma = 10^{-2}$. This demonstrates that
droplet attraction due to spin torque is weak relative to droplet
acceleration due to field inhomogeneity.  A strong enough field
gradient, on the scale of $\alpha \sigma/2\rho_*$, can eject the
droplet from the nanocontact, causing a drift instability previously
observed in numerical simulations \cite{hoefer2010theory}.
Additionally, the associated velocity scale from Eq.~\eqref{eq:14} is
$V = bV_1 \sim \sigma/2$ which is much smaller than $\omega$ as
required for this order of accuracy of the approximate droplet.

\section{Interacting Droplets}
\label{sec:interacting}

An intriguing, indeed defining, aspect of solitary wave dynamics is
their interaction behavior.  Reference \cite{maiden2014attraction}
undertook a numerical investigation of two interacting droplets by
varying droplet parameters and quantifying the properties of the
solution post-interaction.  It was found that the relative phase
difference between the two droplets plays a fundamental role,
controlling whether the interaction is attractive or repulsive. The
attractive interactions studied were strongly nonlinear, hence a
perturbation theory would be insufficient to study the full complement
of observed phenomena.  Nevertheless, we can gain insight into the
nature of the interaction (attractive/repulsive) by studying two
well-separated droplets perturbatively, with the small parameter being
the inverse of the droplet separation.  This approach is well-known
and has been applied successfully to, for example, NLS-type models
\cite{zhu_universal_2007,ablowitz2009asymptotic}.

In full generality, the perturbations arising from this analysis are
complex. However, since the validity of these equations is strongly
dependent on the separation of the two droplets, we only expect these
equations to be valid over short time scales. Hence, we only seek to
describe the initial behavior of two stationary, weakly overlapping
droplets. As the interaction immediately excites propagation of the
two droplets, this will not model the behavior for $t>0$.
Nevertheless, these assumptions make it possible to describe much of
the behavior observed in full numerical simulations
\cite{maiden2014attraction}.  The initial configuration places one
droplet on the left (subscripted $1$) and another droplet (subscripted
$2$) a distance $d$ away along the $x-$axis. We define the relative
phase difference $\Delta \ph = \ph_2 - \ph_1$, which will emerge as an
important quantity in the modulation equations.  Considering the
modulation equations for two weakly interacting droplets with motion in
the $\hat{\mathbf{x}}$ direction \textit{at the initial time only} yields
\begin{align}
  \label{eq:11}
  \ds\dot{\ph}_{0,k}
    &\ds 
    = -\frac{\w}{2 \pi} \cos(\Delta \ph)\intrr{\mathcal{K}_k(\mathbf{x})},
    \\
    \label{eq:13}
    \ds 
    \dot{\xi}_{k}
    &
    \ds 
    = \frac{\w}{2 \pi } (-1)^{k+1}\sin(\Delta
    \ph)\intrr{\mathcal{K}_k(\mathbf{x}) \sech\left(\rho -\frac{1}{\w}
      \right) \cos \varphi} ,
    \\[2mm]
    \label{eq:12}
    \ds 
    \dot{\omega}_{k}
    & 
    \ds 
    = -\frac{ \omega ^3}{4 \pi } (-1)^{k+1}\sin(\Delta \ph)
    \intrr{\mathcal{K}_k(\mathbf{x})\sech\left(\rho -\frac{1}{\w}
      \right)} ,
    \\[2mm]
  \label{eq:mod_rhs}
    \ds 
    \dot{V}_{k}
    &
    \ds 
    =  \frac{\omega ^3}{ \pi } \cos(\Delta \ph)
    \intrr{\mathcal{K}_k(\mathbf{x}) \cos \varphi},
\end{align}
where 
\begin{align}
  \mathcal{K}_k(\mathbf{x})& =\sech\left(\tilde{\rho}_k -
    \frac{1}{\w}\right)\sech\left(\rho - \frac{1}{\w}\right)  
   \times \left[2\sech^2\left(\rho - \frac{1}{\w}\right) - \w
    \left (1- \tanh\left(\rho - \frac{1}{\w}\right) \right)\right].
  \label{eq:intereacting_integral_term}
\end{align}
The integration kernel $\mathcal{K}_k$ depends on the separation
between the two droplets through $\tilde{\rho}_k= \sqrt{(x+(-1)^k d)^2
  + y^2}$.  Utilizing this framework, we can now offer some insight
into the nature of two interacting droplets.

\subsection{Attraction and Repulsion}
\label{sec:attraction-repulsion}

The attractive or repulsive nature of two droplets can be understood
by considering Eq.~\ref{eq:mod_rhs}.
As $\Delta \ph$ varies, the sign of $\cos(\Delta \ph)$ is
clear. Determining the initial direction of motion, right or left, of
the droplet comes down to determining the sign of the integral term in
\eqref{eq:mod_rhs}.
\begin{figure}[htbp]
  \centering
  \begin{tabular}{cc}
    \includegraphics{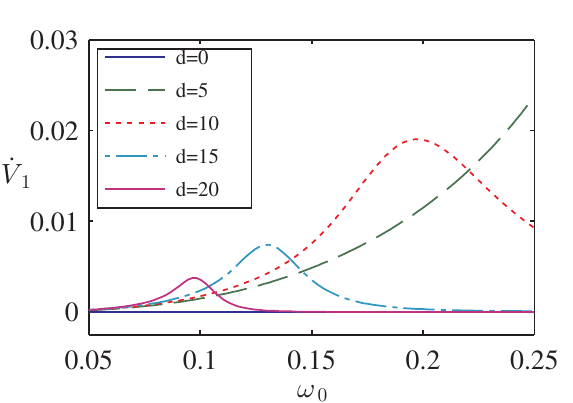} &  \includegraphics{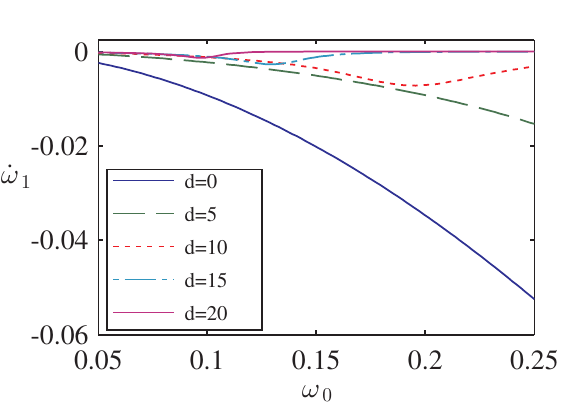}
  \end{tabular}
  \caption{ Left: Initial acceleration for varied initial $\w$ and
    several values of separation.  Right: Plot of $\dot{\w}_1$ as
    a function of initial $\w_0$ for several values of separation.  In
    both, the initial relative phase was $\Delta \ph =1$. }
    \label{fig:dvdt} 
\end{figure}
Figure \ref{fig:dvdt}, left shows the numerical evaluation of the
right hand side of $\dot{V}_1$ (droplet on left) when $\Delta \Phi = 1
< \pi/2$, leading to positive values only. Thus, the left droplet
experiences a positive acceleration to the right, towards the other
droplet when $|\Delta \Phi| < \pi/2$.  Since the kernel exhibits the
symmetry with respect to droplet choice $\mathcal{K}_1(x,y) =
\mathcal{K}_2(-x,y)$, the integral in \eqref{eq:mod_rhs} for the right
droplet, $k = 2$, has the opposite sign.  The right droplet
experiences a negative acceleration to the left when $|\Delta \Phi| <
\pi/2$.  Therefore, two droplets are attractive when $|\Delta \Phi| <
\pi/2$, i.e., when they are sufficiently in phase.  Similarly, when
$\pi/2 < |\Delta \ph| < \pi$, the signs of $\dot{V}_k$ are reversed
and the droplets move away from each other.  Thus, two droplets are
repulsive when they are sufficiently out of phase.

As was noted in \cite{maiden2014attraction} by a nonlinear method of
images, the attractive or repulsive nature of two droplets with the
special initial values $\Delta \Phi = 0$ or $\Delta \Phi = \pi$
describes the behavior of a single droplet near a magnetic boundary
with either a free spin (Neumann type) boundary condition or a fixed
spin (Dirichlet type) boundary condition, respectively.  The analysis
presented here confirms this fact for any droplet that weakly
interacts with a magnetic boundary.  Such behavior was observed in
micromagnetic simulations of a droplet in a NC-STO, nanowire geometry
\cite{iacocca_confined_2014}.

\subsection{Asymmetry}

Despite a highly symmetric initial condition, an asymmetry was
observed in so-called ``head-on collisions" of two droplets in
\cite{maiden2014attraction}.  The frequency equation \eqref{eq:12}
provides an explanation of this in the limit of very small velocities.
Figure \ref{fig:dvdt}, right contains the relevant information. Since
$\dot{w}_1$ appears to be negative always when $\sin(\Delta
\ph)>0$. In the numerical experiments in \cite{maiden2014attraction}
were done over the range $\Delta \ph =0$ to $\Delta \ph = \pi$, this
was always the case. Again using that $\mathcal{K}_1(x,y) =
\mathcal{K}_2(-x,y)$, it can be seen that the integrals involved in
computing $\dot{\w}_1$ and $\dot{\w}_2$ are equal. Hence the sign of
$\dot{\w}_k$ is determined by $(-1)^{k+1}$, and the signs of
$\dot{\w}_1$ and $\dot{\w}_2$ will always be opposite. For the
parameters discussed here, this means that the frequency decreases for
the droplet on the left and increases on the right.  This change in
droplet structure is asymmetric because a reduced (increased)
frequency implies larger (smaller) droplet mass and corresponds
precisely with the observations of \cite{maiden2014attraction}.

\subsection{Acceleration}

\begin{figure}[htbp]
  \centering
  \begin{tabular}{c}
    \includegraphics[width =.46\textwidth]{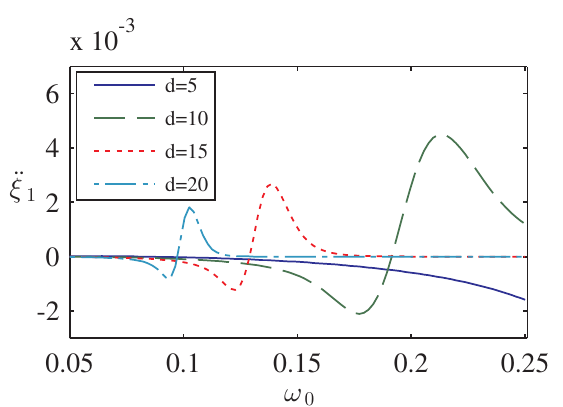}
  \end{tabular}
  \caption{Numerical evaluation of $\ddot{\xi}_1$ initially for
    $\Delta \Phi = \pi/2$, variable droplet separation $d$ and
    frequency $\omega_0$. There is not one sign of acceleration, i.e.,
    the left droplet can be repelled or attracted to the right droplet
    depending on the choice of parameters. }
  \label{fig:acc_plots}
\end{figure}

The discussion of attraction and repulsion in Section
\ref{sec:interacting}\ref{sec:attraction-repulsion} suggests that the
boundary between the two behaviors is $\Delta \ph = \pi/2$.  But this
does not agree with numerical experiments where the crossover $\Delta
\ph$ was found to vary with the initial droplet parameters
\cite{maiden2014attraction}.  To offer an explanation for this, we
consider the total acceleration of the initial droplets, i.e.,
$\ddot{\xi}_k$.  This incorporates higher order information not
included in $\dot{V}_k$. Since the full modulation equations for
interacting droplets when $V \ne 0$ are complex, we do not examine
$\ddot{\xi}_k$ for all values of $\Delta \ph$.
However, at $\Delta \ph = \frac{\pi}{2}$, we know
$\dot{V}_k=\dot{\ph}_0=0$, (since $\cos(\Delta \ph) = 0$) and those
terms will not contribute, which simplifies the calculation.  Figure
\ref{fig:acc_plots} shows the initial, total droplet acceleration
$\ddot{\xi}_1$, evaluated numerically, as the initial frequency and
separation are varied.
The variable sign of this quantity as parameters change demonstrates
that subtle, higher order effects causes the crossover value of
$\Delta \ph$ to deviate from its nominal value $\pi/2$.

\section{Conclusion}
The primary contribution of this work is a general framework for
investigating perturbations of droplet solitons.  Actual physical
devices used to create and manipulate droplets are quite complex,
incorporating a number of physical effects.  Therefore, having a
tractable, analytical theory to describe both the motion and
precession of droplets due to physical perturbations is quite
valuable.
The examples presented here are meant to demonstrate the versatility
and power of this tool.  Additionally, the application of this theory
to the NC-STO provides several insights into the behavior of
experimentally observed dissipative droplets. In particular, the
dissipative droplet is shown to be robust in the presence of weak
field gradients, but can be ejected from the nanocontact if the field
gradient is too large, providing an explanation for a previously
observed drift instability.  These observations open possible
mechanisms for generating a current of solitons which could serve as a
mechanism for information transfer.

As demonstrated by examples in the preceding sections, many
perturbations excite evolution of higher order parameters (overall
phase and position).  This subtle information proves to be of
fundamental importance for several perturbations considered.  As shown
by our derivation of the modulation equations for a general class of
Hamiltonian systems, the higher order parameter dynamics emerge when
the generalized nullspace of the linearized evolution operator is
incorporated.  Due to the existence of an approximate analytical form
for the propagating droplet, we are able to completely characterize
this nullspace and hence recover the droplet modulation equations in a
convenient form.

A number of physical perturbations can now be investigated within this
framework.  Future developments of the modulation theory itself could
be performed in the context of a different family of magnetic droplet
solutions.  One example is the weakly nonlinear droplet
\cite{ivanov2001small}.  Recent micromagnetic simulations have found
rotating and precessing localized waves in NC-STOs with non-trivial
magnetostatic contributions \cite{finocchio_nanoscale_2013}.  The
invariance of equation \eqref{eq:nondimensional_ll} when $\e = 0$ with
respect to rotation of the domain and an analysis of conserved
quantities \cite{papanicolaou1991dynamics}, suggests that there may be
rotating and precessing solitary wave solutions.  The modulation
theory developed in this work could be extended to such solitary wave
solutions.

\section*{Acknowledgment}
The authors gratefully acknowledge support through an
NSF CAREER grant.

\bibliographystyle{plainnat} 
\bibliography{Moving_Droplet_PRSA}
\appendix
\section{Numerical Method}
\label{app:mu-magnetics}

The numerical simulations (micromagnetics) we conducted incorporated a
periodic Fourier psuedospectral spatial discretization.  Unless
otherwise stated, the spatial domain was $[-50,50]\times[-50,50]$,
sufficiently large so that the perturbed solitary waves were
well-localized within it.  In each spatial dimension, $2^9$ grid
points were used. The time-stepping was done using a version of the
Runge-Kutta-Fehlberg algorithm, modified so that the magnetization
maintained unit length at every grid point and each time step.

The velocity, $\mathbf{V}$, was extracted from numerical data by
computing the center of mass, $\boldsymbol{\xi}(t) =
\int_{\mathbb{R}^2} \mathbf{x} (1 - m_3(\mathbf{x},t))
d\mathbf{x}/\mathcal{N}$. 
We then estimated $\mathbf{V} = \dot{\boldsymbol{\xi}}$, approximated
using a forward difference of $\boldsymbol{\xi}(t)$. This method does not work for perturbations which excite higher order changes in $\dot{\boldsymbol{\xi}}$ and we do not estimate $\mathbf{V}$ in such cases.

 For the
precessional frequency $\omega$, the phase of the in-plane
magnetization $(m_x,m_y)$ was extracted at a point a fixed distance
from the center of mass $\boldsymbol{\xi}$.  Differentiating this
phase with respect to time yields $\Omega(t)$ as in
Eq.~\eqref{eq:5}. The precessional frequency, $\w$, was obtained by
subtracting $h_0$ and $\dot{\Phi}_0$. The contribution from
$\dot{\Phi}_0$ was estimated via the modulation equation
\eqref{eq:gen_ph_eqn}. An alternative method based on computing the
conserved quantities in
Eqs.~\eqref{eq:total_spin_def}-\eqref{eq:momentum_def} was used for
comparison. The relations for total spin and momentum of the
approximate droplet
Eqs.~\eqref{eq:total_spin_approx}-\eqref{eq:momentum_approx} were
inverted to obtain $\w$ and $\mathbf{V}$. The two methods were in good
agreement.

\section{Approximate Droplet}
\label{app:approximate_droplet}
As noted in the main body of the manuscript, the derivation of the
approximate droplet is significantly simplified by exploiting the
invariance of Eq. \eqref{eq:droplet_pde} under rotation of the domain
and working in the frame where $V_y= 0$ and $\mathbf{V} = V_x
\hat{\mathbf{x}} \equiv V \hat{\mathbf{x}}$. The derivation proceeds by
substituting the ansatz
\begin{equation} 
  \Th = \Th_0(\rho) + V \Th_1(\rho,\varphi) + \bo{V^2} \mbox{ and }
  \ph = \ph_0 +   V \ph_1(\rho,\varphi) + \bo{V^2} 
  \label{eq:approx_droplet_ansatz}
\end{equation}
into Eq. \eqref{eq:droplet_pde}. At order \bo{1}, this yields one
nontrivial equation,
\begin{equation}
  \frac{d^2\Th_0}{d \rho^2} + \frac{1}{\rho} \frac{d\Th_0}{d \rho} +
  \left(\w -\cos \right) \sin(\Th_0) = 0 .
  \label{eq:approx_droplet_order_one}
\end{equation}
This equation has been extensively studied and is known to admit an
approximate solution in the limit of $0< \w \ll 1$, $\Th_0 =
\arccos\left( \tanh\left( \rho - \frac{1}{\w} \right)\right) +
\bo{\w^2}$
\cite{ivanov1989two,kosevich__1986,bookman2013analytical}.  From
here on out, we additionally assume that $0<\w\ll1$. At order
$\bo{V}$, we find
\begin{align}
  \label{eq:approx_droplet_order_V_nontrivial}
  \sin(\Th_0)\Delta \ph_1  + \left( \cos(\varphi) + 2 \cos(\Th_0)
    \frac{\partial \ph_1}{\partial \rho}\right) \frac{ d \Th_0}{d
    \rho}= 0 ,\\
  \label{eq:approx_droplet_order_V_trivial}
  \Delta \Theta _1 + \left(\omega  \cos \left(\Theta _0\right)-\cos \left(2 \Theta
      _0\right)\right) \Theta _1 = 0. 
\end{align}
Equation \eqref{eq:approx_droplet_order_V_trivial} is solved by
$\Theta_1 = 0$. Substituting the approximate solution for $\Th_0$ into
Eq. \eqref{eq:approx_droplet_order_V_nontrivial} 
\begin{equation}
  \left(\Delta \ph_1  - \cos(\varphi) - 2  \tanh\left( \rho -
      \frac{1}{\w}\right) \frac{\partial \ph_1}{\partial \rho}\right)
  \sech\left( \rho - \frac{1}{\w}\right) = 0 
  \label{eq:approx_droplet_order_V_nontrivial_first_reduction}
\end{equation} 
The residual in
Eq.~\eqref{eq:approx_droplet_order_V_nontrivial_first_reduction} is
determined by two considerations.  If $1 \ll \rho - \w^{-1}$,
$\sech\left( \rho - 1/\w\right)$ dominates and the residual is
exponentially small. In the other case, i.e.  $\rho \sim \w^{-1}$, the
residual will only be small if $\Delta \ph_1 - \cos(\varphi) - 2
\tanh\left( \rho - 1/\w\right) \frac{\partial \ph_1}{\partial \rho}$
is small since $\sech\left( \rho - 1/\w\right)$ is $\bo{1}$. This
suggests that the boundary condition, $\lim_{\rho} \grad \ph_1 =
\frac{1}{2}\hat{\mathbf{x}}$, may be neglected.
Assuming $\ph_1$ is separable of the form $\ph_1(\rho,\varphi) =
f(\rho) \cos(\varphi)$,
Eq. \eqref{eq:approx_droplet_order_V_nontrivial_first_reduction}
simplifies to the ordinary differential equation
\begin{equation}
  \frac{d^2 f}{d \rho^2} + \left(\frac{1}{\rho} - 2 \tanh\left(\rho -
      \frac{1}{\omega}\right) \right)\frac{df}{d \rho} -
  \frac{1}{\rho^2} f = 1 . 
  \label{eq:phi_V_ode}
\end{equation}
Numerical solutions of Eq. \eqref{eq:phi_V_ode} demonstrate that $f$
becomes quite large, approximately $\bo{\frac{1}{\w^2}}$ near $\rho =
\frac{1}{\omega}$. Factoring this into the analysis, we change to the
coordinate system $R = \rho -\frac{1}{\omega}$ and expand $f$ in the
series
\begin{equation}
  f(\rho) = \frac{f_0(\rho)}{\omega^2} + \frac{ f_1(\rho)}{\omega} +
  f_2(\rho) + \cdots . 
  \label{eq:f_ansatz}
\end{equation} 
Let $L \equiv \frac{d^2}{d R^2} - 2 \tanh(R) \frac{d}{d
  R}$. Substituting the ansatz in Eq. \eqref{eq:f_ansatz} into
Eq. \eqref{eq:phi_V_ode} yields,
\begin{align}
\bo{\frac{1}{\w^2}} : \hspace{-2.5cm} && L f_0 &= 0 \label{eq:phi_v_neg2}\\
\bo{\frac{1}{\w}} : \hspace{-2.5cm} &&   L f_1 &= -\frac{d f_0}{d R}\label{eq:phi_v_neg1}\\
\bo{1} : \hspace{-2.5cm} && L f_2 &=  -\frac{d f_1}{d R}  + 1 +f_0 +R \frac{d f_0}{d R}\label{eq:phi_v_1}
\end{align}
Eq. \eqref{eq:phi_v_neg2} admits any constant solution. Take $f_0 =
A$. Substituting this expression for $f_0$ into
Eq. \eqref{eq:phi_v_neg1}, yields $Lf_1 = 0$. Thus, any constant
solution is admissible for $f_1$ as well. Take $f_1 = B$.
Substituting these expressions for $f_0$ and $f_1$ into
Eq. \eqref{eq:phi_v_1}, yields $L f_2 = 1 +A$. However, constant
solutions are in the kernel of $L$, so solvability of this equation
requires $A=-1$.  Similarly, solvability at $\bo{\w}$ requires
$B=0$. Hence, we take $f(\rho) = -\frac{1}{\w^2}$, which
gives rise to the form of the approximate droplet in
Eq. \eqref{eq:approxphi}. 
 
Since $\Phi$ enters into Eq. \ref{eq:droplet_pde} only in the form of $\grad \Phi$, the asymptotic approximations here are valid provided that $\grad{\Phi}$ remains small. This condition is not equivalent  $V \Phi_1 \ll \Phi_0$. Given the form of the approximation for $f$, the small gradient condition requires that $ \abs{\grad \Phi} =\abs{ \frac{V}{\w^2}  \frac{\sin \varphi }{ \rho} }\ll1$. Since the approximation is localized at $\rho\sim\w^{-1}$, we require $V \ll \w$. 
 \end{document}